\documentclass[12pt]{article}
\usepackage{epsfig}
\usepackage{cite}

\def\beq{\begin{equation}}
\def\eeq{\end{equation}}
\def\beqa{\begin{eqnarray}}
\def\eeqa{\end{eqnarray}}
 
\newlength{\dinwidth} \newlength{\dinmargin}
\begin{document}

\begin{center}
{\Large \bf Soft anomalous dimensions for single-top production \newline at three loops}
\end{center}
\vspace{2mm}
\begin{center}
{\large Nikolaos Kidonakis}\\
\vspace{2mm}
{\it Department of Physics, Kennesaw State University,\\
Kennesaw, GA 30144, USA}
\end{center}
 
\begin{abstract}
I present results for soft anomalous dimensions in single top-quark production processes through three loops. I first  
discuss the cusp anomalous dimension with one massive and one massless line at three loops. 
I then calculate the three-loop soft anomalous dimension for $tW$ production and for related processes with new physics
such as $tZ$, $t\gamma$, and $tH^-$ production. This is followed by results at two and three loops for 
the soft anomalous dimension matrix in $s$-channel single-top production and, finally, in $t$-channel single-top production.
These new results are needed for soft-gluon resummation at N$^3$LL accuracy and for calculations of soft-gluon corrections at N$^3$LO.
\end{abstract}
 
\section{Introduction}

Calculations of higher-order corrections to scattering cross sections are central to the exploration of particle 
physics at hadron colliders. The understanding of the structure of these cross sections can be deepened by the 
study of renormalization-group properties of the functions that describe the emission of quanta in the partonic processes. 
In particular, the study of soft and collinear gluon emission is important not only in exploring the rich features of quantum field theory but also in extending the calculations to higher orders. The study of soft-gluon corrections is important and relevant for top-quark production, which is a central topic in collider physics  (see Ref. \cite{NKtoprev} for a recent review). 

Soft and collinear gluon corrections can be resummed to all orders in perturbation theory. Such considerations start with the factorization properties of the cross section. We consider single-top partonic processes of general form $f_{1} + f_{2} \rightarrow t+ X$. The incoming and outgoing particles are represented by eikonal (Wilson) lines, which are gauge-field path-ordered exponentials in the direction of their four-velocity. 

We can define moments of the partonic cross section by 
${\hat\sigma}(N)=\int (ds_4/s) \,  e^{-N s_4/s} \, {\hat\sigma}(s_4)$, where 
$N$ is the moment variable and $s_4$ is a kinematical variable that indicates distance from partonic threshold, i.e. $s_4$ measures the invariant mass squared of additional final-state radiation and it vanishes at threshold. The factorized cross section in moment space takes the form \cite{NKGS}
\beq
{\hat \sigma}^{f_1 f_2\rightarrow tX}(N)=
H_{IL}^{f_1 f_2\rightarrow tX}(\mu) \, 
S_{LI}^{f_1 f_2 \rightarrow tX}\left(N, \mu\right)  
\prod  J_{\rm in} \left(N,\mu \right)
J_{\rm out} \left(N,\mu \right)  \, .
\label{factor}
\eeq
The function $H$ is a short-distance hard function and does not depend on
$N$. The function $S$ is a soft function that encodes 
the coupling of soft gluons to the partons in the
scattering. These process-dependent $H$ and $S$ functions 
are in general matrices in color space for the partonic scattering, 
and $I$ and $L$ are indices that label color tensors. 
The functions $J_{\rm in}$ and $J_{\rm out}$, on the other hand, 
describe universal soft and collinear emission from incoming and outgoing  
partons, respectively. The functions $S$, $J_{\rm in}$, and $J_{\rm out}$, depend 
on the moment variable $N$ as well as on the scale $\mu$.

The soft function, $S$, obeys the renormalization-group equation \cite{NKGS}
\beq
\left(\mu \frac{\partial}{\partial \mu}
+\beta \frac{\partial}{\partial g_s}\right)\,S_{LI}
=-(\Gamma^\dagger_S)_{LK}S_{KI}-S_{LK}(\Gamma_S)_{KI}
\eeq
where $g_s^2=4\pi\alpha_s$, $\alpha_s$ is the strong coupling, $\beta$ is the QCD beta function,
$\beta=\mu^2 d\ln\alpha_s/d\mu^2$, and $\Gamma_S$ is the 
process-dependent soft anomalous dimension, a matrix in color space.

The resummed cross section \cite{NKGS} is derived from the renormalization-group evolution of the functions in Eq. (\ref{factor}), and it takes the form of a product of exponentials. The exponents that resum universal soft and collinear radiation from incoming and outgoing partons were derived in \cite{GS,CT}, while noncollinear soft radiation depends on the color structure of the particular process and is resummed via exponentials of $\Gamma_S$ \cite{NKGS}.

Apart from their intrinsic theoretical value, the soft-gluon corrections have turned out to be extremely important for many processes in hadron colliders, in particular the Tevatron and the LHC.
Next-to-leading-order (NLO) calculations are now standard for processes studied at the LHC, and next-to-next-to-leading-order (NNLO) corrections are the new frontier. In some cases
even next-to-next-to-next-to-leading-order (N$^3$LO) corrections may be necessary to meet the increasing experimental precision.  
The soft-gluon contributions tend to dominate the higher-order corrections for many processes, not only near partonic threshold but even relatively far from it. Thus, calculations that involve soft-gluon corrections
are usually very good approximations to exact results, and they can be used to produce excellent approximate predictions at higher orders.
Results are known for a multitude of processes including top quarks, Higgs and electroweak bosons, etc. (see e.g. \cite{NKtoprev} for a review).
Approximate NNLO (aNNLO) and approximate N$^3$LO (aN$^3$LO) calculations that include soft-gluon corrections provide solid predictions for all top-quark production processes that have been studied.

We first discuss the cusp anomalous dimension, which is the simplest soft anomalous dimension, with one massive and one massless eikonal line through three loops. We then present results for the associated production of a top quark with a $W$ boson and related processes. Finally, we present results for $s$-channel single-top production and for $t$-channel single-top production. These are the first three-loop calculations of soft anomalous dimensions for processes with one massive quark.

\section{Cusp anomalous dimension with one massive quark}

The cusp anomalous dimension is a fundamental quantity in perturbative QCD, and it describes the infrared behavior of scattering amplitudes \cite{AMP,BNS,IKR,KR,NK2loop,CHMS,HH,GHKM,NK3loop}. The cusp angle $\theta$ of two eikonal lines with momenta $p_i$ and $p_j$ is given by 
$\theta=\cosh^{-1}(p_i\cdot p_j/\sqrt{p_i^2 p_j^2})$.

The perturbative series for the cusp anomalous dimension in QCD is written as 
\beq
\Gamma_c=\sum_{n=1}^{\infty} \left(\frac{\alpha_s}{\pi}\right)^n \Gamma^{(n)}_c \, .
\eeq 
The $n$th-order cusp anomalous dimension, $\Gamma^{(n)}_c$, is determined from the ultraviolet poles in dimensional regularization of $n$-loop diagrams calculated in the eikonal approximation.

When both eikonal lines are massive, the expressions get very complicated at higher orders. This is in contrast to the light-like cusp anomalous dimension, where the results are proportional to the cusp angle, with proportionality factor denoted as $K^{(n)}$ at $n$ loops.
In the massive case the expressions involve nonelementary functions of the cusp angle, including (harmonic) polylogarithms. However, the expressions are simpler when only one of the quarks has nonzero mass. 

\begin{figure}
\begin{center}
\includegraphics[width=10cm]{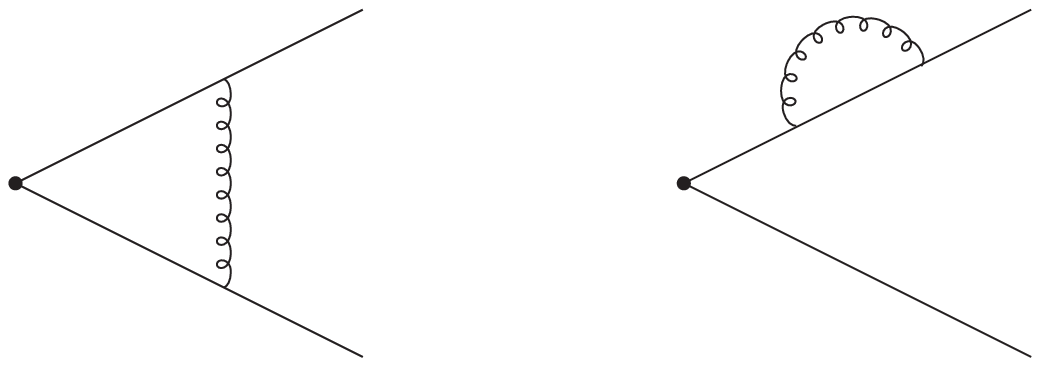}
\caption{One-loop diagrams for the cusp anomalous dimension.}
\label{loop1plot}
\end{center}
\end{figure}

The one-loop diagrams for the cusp anomalous dimension are shown in Fig. \ref{loop1plot}. Assuming that eikonal line $i$ represents a massive quark and eikonal line $j$ represents a massless quark, we find at one loop 
\beq
\Gamma_c^{(1)}=C_F \left[\ln\left(\frac{2 p_i \cdot p_j}{m_i \sqrt{s}}\right) -\frac{1}{2}\right] 
\label{1loop1}
\eeq  
where $m_i$ is the mass of the quark in eikonal line $i$, $s=(p_i+p_j)^2$ and $C_F=(N_c^2-1)/(2N_c)$, with $N_c=3$ the number of colors.

\begin{figure}
\begin{center}
\includegraphics[width=10cm]{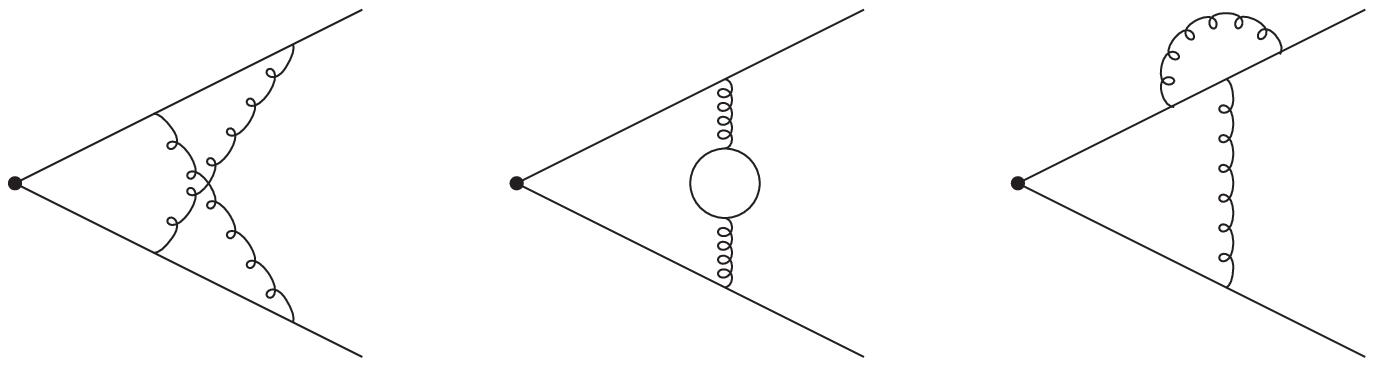}
\caption{Representative two-loop diagrams for the cusp anomalous dimension.}
\label{loop2plot}
\end{center}
\end{figure}

The two-loop cusp anomalous dimension was studied in \cite{KR} and a result was provided in terms of a few uncalculated integrals. A later independent calculation provided fully analytical explicit results in terms of dilogarithms and trilogarithms that were presented in \cite{NK2loop}. Some representative two-loop diagrams for the cusp anomalous dimension are shown in Fig. \ref{loop2plot}.

When one of the quarks (with momentum $p_j$) is massless the expression is simpler, and we find at two loops 
\beq
\Gamma^{(2)}_c=K^{(2)} \left[\ln\left(\frac{2 p_i \cdot p_j}{m_i \sqrt{s}}\right) -\frac{1}{2} \right]+\frac{1}{4} C_F C_A (1-\zeta_3)
\label{2loop1}
\eeq 
where $K^{(2)}=C_F C_A (67/36-\zeta_2/2)-5 T_F n_f/9$ with $\zeta_2=\pi^2/6$, $C_A=N_c$, $T_F=1/2$, and $n_f$ is the number of light quark flavors (for top production $n_f=5$).

\begin{figure}
\begin{center}
\includegraphics[width=10cm]{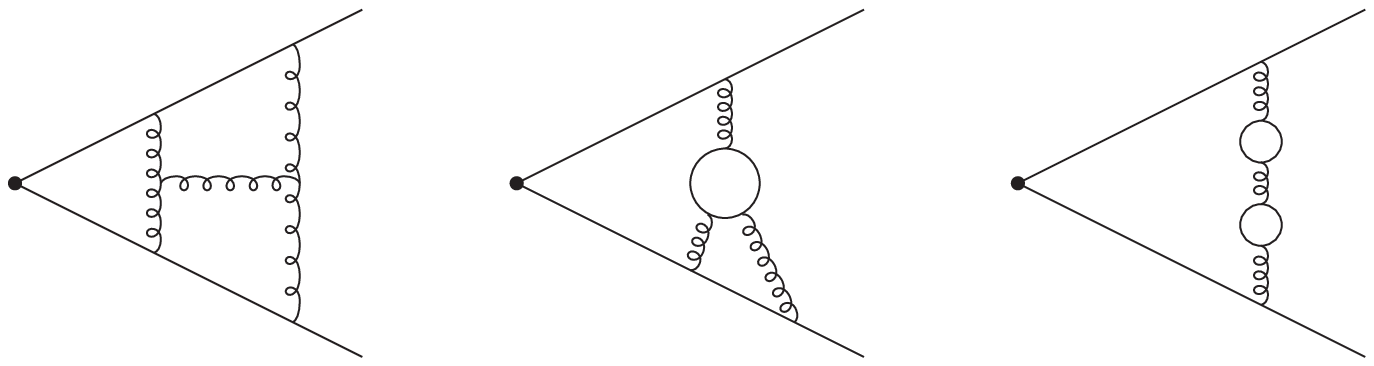}
\caption{Representative three-loop diagrams for the cusp anomalous dimension.}
\label{loop3plot}
\end{center}
\end{figure}

The three-loop cusp anomalous dimension was calculated in Ref. \cite{GHKM} and presented in terms of a large number of harmonic polylogarithms, defined iteratively with up to five integrations. Those results were reexpressed in terms of ordinary polylogarithms and a few simple integrals of them in Ref. \cite{NK3loop}, and simple approximations were also presented. A few representative three-loop diagrams for the cusp anomalous dimension are shown in Fig. \ref{loop3plot}. 

Again, we find that the result is simpler and shorter when one quark ($p_j$) is massless. At three loops, our calculation gives the expression
\beq
\Gamma^{(3)}_c=K^{(3)} \left[\ln\left(\frac{2 p_i \cdot p_j}{m_i \sqrt{s}}\right)-\frac{1}{2}\right]+ \frac{1}{2}K^{(2)} C_A (1-\zeta_3)+C^{(3)}
\label{3loop1}
\eeq
where
\beqa
K^{(3)}&=&C_F C_A^2 \left(\frac{245}{96}-\frac{67}{36} \zeta_2
+\frac{11}{24}\zeta_3+\frac{11}{8}\zeta_4\right)
+C_F C_A T_F n_f \left(-\frac{209}{216}+\frac{5}{9}\zeta_2
-\frac{7}{6}\zeta_3\right)
\nonumber \\ &&
+C_F^2 T_F n_f \left(\zeta_3-\frac{55}{48}\right)
-\frac{1}{27} C_F T_F^2 n_f^2 
\label{K3}
\eeqa
with $\zeta_3 \approx 1.202$ and $\zeta_4=\pi^4/90$,
and
\beq
C^{(3)}=C_F C_A^2\left[-\frac{1}{4}+\frac{3}{8}\zeta_2-\frac{\zeta_3}{8}-\frac{3}{8}\zeta_2 \zeta_3+\frac{9}{16} \zeta_5\right]
\label{C3}
\eeq
with $\zeta_5 \approx 1.0369$. The above result involves taking the limit of the explicit three-loop expression in \cite{NK3loop} as the mass of one parton goes to zero, properly taking into account self-energies (as we also did for the one- and two-loop results).

The cusp anomalous dimension is a necessary ingredient of any calculation of soft anomalous dimensions in $2 \rightarrow m$ multi-leg scattering processes.
When all external lines are massless, it was shown in \cite{ADS} that the two-loop soft anomalous dimension matrix is proportional to the one-loop result.
Further studies of the structure of soft anomalous dimensions with massless partons were presented in \cite{BN,GM,LJD}. Three-parton correlations do not contribute to the soft anomalous dimension with massless partons at any order due to scaling symmetry constraints \cite{BN,GM}. In Ref. \cite{ADG} three-loop results for the soft anomalous dimension in massless multi-leg scattering were presented, including four-parton correlations. When both of the eikonal lines are massive then the three-parton correlations no longer vanish \cite{FNPY}; however, they still vanish when only one of the lines is massive.

We use our results for the cusp anomalous dimension as an ingredient in our calculations below of the soft anomalous dimensions for single-top production processes, including $tW$, $s$-channel and $t$-channel production \cite{NKsintop,NKtW,NKs,NKt}, as well as related processes in new-physics models. These are the first such calculations at three loops for processes with one massive quark.

\section{$tW$ production and related processes}

We now calculate the three-loop soft anomalous dimension for the associated production of a top quark with a $W$ boson
via $bg\rightarrow tW^-$. The cross section for this process is known at NLO \cite{Zhu} and with additional higher-order soft-gluon corrections \cite{NKsintop,NKtW}, and it has most recently been measured in proton-proton collisions at the LHC at 13 TeV energy \cite{ATLAStW13,CMStW13}. 

The soft anomalous dimension for $tW$ production is identical to that  
for other related processes in models of new physics, such as the associated production of a top quark with a charged Higgs boson 
via $bg\rightarrow tH^-$ in two-Higgs-doublet models \cite{NKtH}; the associated production of a top quark with a $Z$ boson with anomalous 
top-quark couplings via $qg\rightarrow tZ$ \cite{NKtZgam,NKtZ}; and the associated production of a top quark with a photon with anomalous top-quark 
couplings via $qg\rightarrow t\gamma$ \cite{NKtZgam,FKtgam}. In all these cases we have $2 \rightarrow 2$ processes at lowest order that involve two massless partons, a final-state top quark, and a final-state boson. Thus, we have three colored particles involved in the scattering, one of which is massive.

We consider $tW$ production via the process
\beq
b(p_1)+g(p_2) \rightarrow t(p_3) +W^-(p_4)
\eeq
and define $s=(p_1+p_2)^2$, $t=(p_1-p_3)^2$, $u=(p_2-p_3)^2$. The threshold variable is $s_4=s+t+u-m_t^2-m_W^2$, where $m_t$ is the top-quark mass and $m_W$ is the $W$-boson mass.

The soft anomalous dimension for $bg \rightarrow tW^-$ was calculated at one loop in Refs. \cite{NKsintop} and \cite{NKtW}. The one-loop result is given by 
\beq
\Gamma_S^{tW \, (1)}=C_F \left[\ln\left(\frac{-t+m_t^2}{m_t\sqrt{s}}\right)
-\frac{1}{2}\right] +\frac{C_A}{2} \ln\left(\frac{u-m_t^2}{t-m_t^2}\right) \, .
\eeq
This one-loop result is needed for next-to-leading-logarithm (NLL) resummation, 
and it starts contributing to the soft-gluon corrections at NLO with a term $(\alpha_s/\pi) \, 2 \, \Gamma_S^{tW \, (1)} [1/s_4]_+$ times the leading-order factor.

The two-loop result was presented in \cite{NKtW} and is given by
\beq
\Gamma_S^{tW \, (2)}=K^{'(2)} \Gamma_S^{tW \, (1)}
+\frac{1}{4}C_F C_A (1-\zeta_3) 
\eeq
where $K^{'(2)} =K^{(2)} /C_F$.
This two-loop result is needed for next-to-next-to-leading-logarithm (NNLL) 
resummation, and it contributes to the soft-gluon corrections starting at NNLO 
with a term $(\alpha_s/\pi)^2 \, 2 \, \Gamma_S^{tW \, (2)} [1/s_4]_+$ times the 
leading-order factor.

We calculate the soft anomalous dimension at three loops, and find
\beq
\Gamma_S^{tW \, (3)}=K^{'(3)} \Gamma_S^{tW \, (1)}+\frac{1}{2} K^{(2)} C_A (1-\zeta_3)+C^{(3)} \, ,
\eeq
where $K^{'(3)}=K^{(3)}/C_F$, with $K^{(3)}$ defined in Eq. (\ref{K3}), and $C^{(3)}$ is defined in Eq. (\ref{C3}).
This three-loop result is needed for next-to-next-to-next-to-leading-logarithm
(N$^3$LL) resummation,  
and it contributes to the soft-gluon corrections starting at N$^3$LO with a term $(\alpha_s/\pi)^3 \, 2 \, \Gamma_S^{tW \, (3)} [1/s_4]_+$ times the leading-order factor.

\section{Single-top $s$-channel production}

We continue with single-top production via $s$-channel processes.
The cross section for this channel is known at NLO \cite{BWHL}, NNLO \cite{NNLOsch}, and with soft-gluon corrections \cite{NKsintop,NKs}, and it has been measured at the LHC at 8 TeV energy \cite{ATLASsch8,CMSsch8}.

The partonic processes are $q(p_1)+{\bar q}'(p_2) \rightarrow t(p_3) +{\bar b}(p_4)$.
The partonic kinematical variables are 
$s=(p_1+p_2)^2$, $t=(p_1-p_3)^2$, and $u=(p_2-p_3)^2$. 
The threshold variable is given by $s_4=s+t+u-m_t^2$.
In this channel we have $2 \rightarrow 2$ processes at lowest order that involve two massless initial-state partons, a final-state top quark, and a final-state massless quark. Thus, we have four colored particles involved in the scattering, one of which is massive.

The color structure of the hard scattering in $s$-channel single-top production is more complicated than that for 
$tW$ production. Thus, the soft anomalous dimension is a $2 \times 2$ matrix in color space (as it also is for $t{\bar t}$ production via the $q{\bar q}$ channel \cite{NKGS,FNPY,NKtt}). 
We choose a singlet-octet $s$-channel color basis, $c_1=\delta_{a_1 a_2} \delta_{a_3 a_4}$ and $c_2=T^c_{a_2 a_1} T^c_{a_3 a_4}$. We note that in this basis, only the first element of the leading-order hard matrix is nonzero, which leads to some simplifications and has implications for the contributions of the various soft anomalous dimension matrix elements, as we will see below.

The four elements of the $s$-channel soft anomalous dimension matrix at one loop are given by 
\beqa
\Gamma_{11}^{s \, (1)}&=&C_F \left[\ln\left(\frac{s-m_t^2}{m_t\sqrt{s}}\right)
-\frac{1}{2}\right] \, ,
\nonumber \\
\Gamma_{12}^{s \, (1)}&=&\frac{C_F}{2N} \ln\left(\frac{t(t-m_t^2)}{u(u-m_t^2)}\right) \, ,
\nonumber \\
\Gamma_{21}^{s \, (1)}&=& \ln\left(\frac{t(t-m_t^2)}{u(u-m_t^2)}\right) \, ,
\nonumber \\
\Gamma_{22}^{s \, (1)}&=&C_F \left[\ln\left(\frac{s-m_t^2}{m_t \sqrt{s}}\right)-\frac{1}{2}\right]
-\frac{1}{N}\ln\left(\frac{t(t-m_t^2)}{u(u-m_t^2)}\right)
+\frac{N}{2} \ln\left(\frac{t(t-m_t^2)}{s(s-m_t^2)}\right) \, .
\label{Gamma1s}
\eeqa
We note that the first element of the matrix, $\Gamma_{11}^{s \, (1)}$, was calculated in Ref. \cite{NKsintop} and \cite{NKs}, and it is needed for NLL resummation. This element
starts contributing to the soft-gluon corrections at NLO with a term $(\alpha_s/\pi) \, 2 \, \Gamma_{11}^{s \, (1)} [1/s_4]_+$ times the leading-order factor.

For NNLL resummation we also need the other matrix elements in Eq. (\ref{Gamma1s}). The off-diagonal one-loop matrix elements, 
$\Gamma_{12}^{s \, (1)}$ and $\Gamma_{21}^{s \, (1)}$, were given in \cite{NKs} and they contribute to the soft-gluon corrections starting at NNLO. The last one-loop matrix element, $\Gamma_{22}^{s \, (1)}$, was presented in \cite{NKtoprev} and starts contributing at N$^3$LO. 

At two loops, our calculation gives
\beqa
\Gamma_{11}^{s \, (2)}&=& K^{'(2)} \Gamma_{11}^{s \, (1)}+\frac{1}{4} C_F C_A (1-\zeta_3) \, ,
\nonumber \\
\Gamma_{12}^{s \, (2)}&=& K^{'(2)} \Gamma_{12}^{s \, (1)} \, ,
\nonumber \\
\Gamma_{21}^{s \, (2)}&=& K^{'(2)} \Gamma_{21}^{s \, (1)} \, ,
\nonumber \\
\Gamma_{22}^{s \, (2)}&=& K^{'(2)} \Gamma_{22}^{s \, (1)}+\frac{1}{4} C_F C_A (1-\zeta_3) \, .
\label{Gamma2s}
\eeqa
We note that the first element of the matrix, $\Gamma_{11}^{s \, (2)}$, was calculated in Ref. \cite{NKs}, it is needed for NNLL resummation, and it contributes to the soft-gluon corrections starting at NNLO with a term $(\alpha_s/\pi)^2 \, 2 \, \Gamma_{11}^{s \, (2)} [1/s_4]_+$ times the leading-order factor. The other two-loop matrix elements in Eq. (\ref{Gamma2s}) are needed for N$^3$LL resummation, and they have not been presented before. The off-diagonal two-loop matrix elements, $\Gamma_{12}^{s \, (2)}$ and $\Gamma_{21}^{s \, (2)}$, contribute to the soft-gluon corrections starting at N$^3$LO. The last two-loop matrix element, $\Gamma_{22}^{s \, (2)}$, contributes beyond N$^3$LO.  

At three loops, we only need the first element of the soft anomalous dimension matrix for N$^3$LL resummation and to calculate the N$^3$LO soft-gluon corrections. 
The reason for this is that only the first element of the leading-order hard matrix is nonzero, and thus the other elements of the three-loop soft anomalous dimension matrix do not contribute to the N$^3$LO corrections.

We calculate the first element of the $s$-channel soft anomalous dimension matrix at three loops, and find
\beq
\Gamma_{11}^{s \, (3)}= K^{'(3)} \Gamma_{11}^{s \, (1)}
+\frac{1}{2} K^{(2)} C_A (1-\zeta_3)  +C^{(3)} \, .
\eeq
We find that four-parton correlations do not contribute to this matrix element due to the simple color structure.
$\Gamma_{11}^{s \, (3)}$ contributes to the soft-gluon corrections starting at N$^3$LO with a term $(\alpha_s/\pi)^3 \, 2 \, \Gamma_{11}^{s \, (3)} [1/s_4]_+$ times the leading-order factor.

\section{Single-top $t$-channel production}

Finally, we discuss single-top production via $t$-channel processes.
The cross section for this channel is known at NLO \cite{BWHL}, NNLO \cite{NNLOtch,BGYZ,BGZ}, and with soft-gluon corrections \cite{NKsintop,NKt}, and it has been recently measured at the LHC at 13 TeV energy \cite{ATLAStch13,CMStch13}.

The partonic processes are $b(p_1)+q(p_2) \rightarrow t(p_3) +q'(p_4)$.
Again, the partonic kinematical variables are 
$s=(p_1+p_2)^2$, $t=(p_1-p_3)^2$, and $u=(p_2-p_3)^2$, and  
the threshold variable is $s_4=s+t+u-m_t^2$.

The color structure of the hard scattering in $t$-channel single-top production is again complicated, and 
the soft anomalous dimension is a $2 \times 2$ matrix in color space. 
We choose a singlet-octet $t$-channel color basis, $c_1=\delta_{a_1 a_3} \delta_{a_2 a_4}$ 
and $c_2=T^c_{a_3 a_1} T^c_{a_4 a_2}$.

The four elements of the $t$-channel soft anomalous dimension matrix at one loop are given by 
\beqa
{\Gamma}_{11}^{t \, (1)}&=&
C_F \left[\ln\left(\frac{t(t-m_t^2)}{m_t s^{3/2}}\right)-\frac{1}{2}\right] \, ,
\nonumber \\
{\Gamma}_{12}^{t \, (1)}&=&\frac{C_F}{2N} \ln\left(\frac{u(u-m_t^2)}{s(s-m_t^2)}\right) \, ,
\nonumber \\
{\Gamma}_{21}^{t \, (1)}&=& \ln\left(\frac{u(u-m_t^2)}{s(s-m_t^2)}\right) \, ,
\nonumber \\
{\Gamma}_{22}^{t \, (1)}&=& C_F \left[\ln\left(\frac{t(t-m_t^2)}{m_t s^{3/2}}\right)-\frac{1}{2}\right]
-\frac{1}{N}\ln\left(\frac{u(u-m_t^2)}{s(s-m_t^2)}\right) 
+\frac{N}{2}\ln\left(\frac{u(u-m_t^2)}{t(t-m_t^2)}\right) \, .
\label{Gamma1t}
\eeqa
The first element of the matrix, ${\Gamma}_{11}^{t \, (1)}$, was calculated in Ref. \cite{NKsintop} and \cite{NKt}, and it is needed for NLL resummation. For NNLL resummation we also need the other matrix elements in Eq. (\ref{Gamma1t}). The off-diagonal matrix elements, ${\Gamma}_{12}^{t \, (1)}$ and ${\Gamma}_{21}^{t \, (1)}$, were given in \cite{NKt}. The last matrix element, ${\Gamma}_{22}^{t \, (1)}$, was presented in \cite{NKtoprev}. 

At two loops,  our calculation gives 
\beqa
\Gamma_{11}^{t \, (2)}&=& K^{'(2)} \Gamma_{11}^{t \, (1)}+\frac{1}{4} C_F C_A (1-\zeta_3) \, ,
\nonumber \\
\Gamma_{12}^{t \, (2)}&=& K^{'(2)} \Gamma_{12}^{t \, (1)} \, ,
\nonumber \\
\Gamma_{21}^{t \, (2)}&=& K^{'(2)} \Gamma_{21}^{t \, (1)} \, ,
\nonumber \\
\Gamma_{22}^{t \, (2)}&=& K^{'(2)} \Gamma_{22}^{t \, (1)}+\frac{1}{4} C_F C_A (1-\zeta_3) \, .
\label{Gamma2t}
\eeqa
We note that the first element of the matrix, $\Gamma_{11}^{t \, (2)}$, was calculated in Ref. \cite{NKt} and is needed for NNLL resummation. The other two-loop matrix elements in Eq. (\ref{Gamma2t}) are needed for N$^3$LL resummation, and they have not been presented before. The structure of the results is similar to that for the $s$ channel.

At three loops, as for the $s$-channel, we only need the first element of the soft anomalous dimension matrix for N$^3$LL resummation, and to calculate the N$^3$LO soft-gluon corrections. The reason, as before, is the simple structure of the leading-order hard matrix, so that the other elements of the $t$-channel three-loop soft anomalous dimension matrix do not contribute.

We calculate the first element of the $t$-channel soft anomalous dimension matrix at three loops, and find 
\beq
\Gamma_{11}^{t \, (3)}= K^{'(3)} \Gamma_{11}^{t \, (1)}
+ \frac{1}{2} K^{(2)} C_A (1-\zeta_3) +C^{(3)} \, .
\eeq
Again, we note that four-parton correlations do not contribute to this matrix element. $\Gamma_{11}^{t \, (3)}$ contributes to the soft-gluon corrections starting at N$^3$LO with a term $(\alpha_s/\pi)^3 \, 2 \, \Gamma_{11}^{t \, (3)} [1/s_4]_+$ times the leading-order factor.

\section{Conclusion}

I have presented three-loop results for the cusp anomalous dimension with one massive and one massless eikonal lines, and for the soft anomalous dimensions in single-top production processes in the Standard Model and beyond. These are the first calculations of soft anomalous dimensions for processes with one heavy-quark line at three loops, and they allow N$^3$LL soft-gluon resummation and the determination of N$^3$LO soft-gluon corrections for such processes. These calculations also open the door to other calculations for similar processes involving bottom and charm quarks in the Standard Model as well as to new-physics processes involving a massive colored particle.

\section*{Acknowledgements}
This material is based upon work supported by the National Science Foundation under Grant No. PHY 1820795.

\end{document}